\documentclass[twocolumn,showpacs,preprintnumbers,amsmath,amssymb,prl]{revtex4}

\usepackage{graphicx}
\usepackage{dcolumn}
\usepackage{bm}

\begin{document}

\title{A DFT+Nonhomogeneous DMFT approach for Finite Systems}
\author{Alamgir~Kabir$^{a}$,\altaffiliation{Corresponding author, e-mail address:akabir@knights.ucf.edu} 
 Volodymyr~Turkowski$^{a,b}$, and Talat~S.~Rahman$^{a,b}$}

\affiliation{$^{a}$ Department of Physics,University of Central Florida, Orlando, FL 32816\\
 $^{b}$ NanoScience and Technology Center,
University of Central Florida, Orlando, FL 32816}

\date{\today}

\begin{abstract}
For reliable and efficient inclusion of electron-electron correlation effects in nanosystems we propose a combined density-functional-theory/nonhomogeneous dynamical-mean-field-theory (DFT + DMFT) approach which employs an approximate
 Iterative Perturbative Theory (IPT)  impurity solver. The validity of the method is demonstrated
by successful examination of the size-dependent magnetic properties of iron nanoparticles containing 11-100 atoms. We show that the DFT+ DMFT solution is in very good agreement with experimental data, in particular it does not lead to the
overestimation of magnetization that is found with the DFT and DFT+U techniques. More importantly, we demonstrate that DFT+DMFT approach can be used for accurate and realistic description of nanosystems containing about hundred atoms.
\end{abstract}

  
\maketitle

{\it Introduction.}--Magnetic nanoparticles are promising ingredients for several modern and future technologies. For example, nanoparticles with controllable magnetic moment may be used as contrast agents in magnetic-resonance imaging techniques, as directable drug carriers, and in cancer therapy by hyperthermia.\cite{kim} Furthermore, magnetic particles with high anisotropic energy are needed for energy-harvesting technologies\cite{Jones} as well as for ultra-high-density recording media.\cite{Henderson} In most cases, these promising new applications call for a microscopic understanding of magnetic and other characteristics of nanoparticles.
As is true for bulk phenomena such as  metal-insulator transitions, heavy-fermion charge transport, colossal magneto-resistance, and high-temperature superconductivity, electron-electron correlations are expected to play a role in determining the magnetic properties of nanoparticles.\cite{Bozovic} It is thus important to extend to nanoparticles theoretical and computational techniques that take such correlations into account.

 From the perspective of basic science, nanomaterials are fascinating precisely because their electrons behave quite differently from the electron-like quasi-particles of typical bulk metals. Interactions between electrons (and between electrons and the lattice) give rise to properties vastly different from those expected from noninteracting band structures, leading to fascinating phenomena attributable to the collective response of the system.  

In theoretical examination of systems in which electron correlations have to be included beyond that offered by density functional theory(DFT), often the DFT+U method, where U is the on-site Coulomb repulsion term, is employed.\cite{2,3}
Since the latter is based on the mean-field approximation, albeit static,  it provides in some cases better agreement with observations than DFT alone.
However, this method has its limitations. In particular, since
it describes only long-range spin and orbital order it fails for paramagnetic insulators.
It may fail even for cases in which correlations are not strong, as in bulk transition metals. Also, for systems in which U is expected to be larger than the kinetic energy, DFT+U leads to wrong prediction of spin-ordering temperatures (see, Ref.~\cite{12} and references therein). Furthermore, the neglect of time fluctuations, implies that the DFT+U approach is not suitable for including dynamical effects such as time-resolved local interactions. In this regard, for 2D and 3D extended systems DMFT itself has proven to be a good approximation, especially at intermediate values of 
Coulomb repulsion.\cite{6} 
A further important step in the analysis of the correlation effects in real materials was taken when
DMFT was incorporated into the DFT scheme \cite{10,11}. This combined method has been successfully applied to study the spectral, optical and magnetic properties of bulk systems.\cite{12,13}  In short, one could argue that:
1) DFT is suitable for weakly-correlated systems ($U<<t$ (the interatomic hopping parameter)); 2) 
DFT+U for very strongly correlated systems ($U>>t$);
3) DMFT+DFT for intermediate strengths of electron correlations. As a matter of fact DMFT+DFT is not limited to just the intermediate region, as it has been found to be appropriate also for the other two regimes.\cite{6} It may thus be considered a universal approach capable of
capturing correlation effects of all possible
strengths in local electron-electron interactions.
The regime with intermediate
 values of Coulomb repulsion is, nevertheless, the most challenging and interesting, particularly for nanosystems as the competition between the kinetic and potential (Coulomb repulsion) energies may lead to a variety of novel phenomena both in the ground and excited states. 

Already after Florens' suggestion of the suitability of DMFT for nanosystems\cite{Florens}, several groups applied it to model systems containing a few to several hundred atoms.\cite{Valli,Jacob,Lin,Zgid,Snoek,Blumer} The concept of local correlations which works well for extended systems was thus shown to be applicable also at the nanoscale.
%Most relevant for our purposes are the relatively few studies directed towards realistic description of %the properties of nanoparticles.\cite{Boukhvalov,Turkowski,Jacob,vturkowski} 
For example, Boukhvalov et al. \cite{Boukhvalov} were able to   reproduce the energy gap in the ${\rm Mn_{4}}$ molecular magnet correctly, contrary to the DFT+U
approach, which underestimates the gap. Similarly, the spectral and the transport properties of two- and ten-atom nickel nanocontacts were studied by Jacob et at.~\cite{Jacob}. Our recently proposed  approach, in which we used the Quantum Monte Carlo(QMC) method as a solver for the Anderson impurity problem proved to be appropriate for examining the magnetic properties of small (2-to 5-atom) iron and iron-platinum clusters\cite{Turkowski, vturkowski}. However there are technical issues with the application of the QMC solver to low temperatures and to systems containing more than a few atoms. Note that although Valli et al. \cite{Valli} and Lin et al \cite{Lin} applied DFT+DMFT to systems containing 110 and 50 atoms, respectively, the description was limited to a single atomic orbital, which may not be adequate for most systems.
Since the main interest in nanoscience is in accurate and realistic description of systems containing tens to hundreds and thousands of atoms, it is extremely desirable to develop a DFT+DMFT methodology for the purpose.  

Here we propose a formalism which allows one to introduce an approximate  nano-DMFT, generalized IPT solution for the impurity problem. IPT, a second order perturbative approximation in U, succeeds in capturing many important effects
in bulk correlated systems, including zero-energy quasiparticle resonance and the upper and lower incoherent Hubbard bands in the electron spectral function\cite {12}. In addition to being faster than most other solvers, it also gives results in good agreement with the more accurate QMC and exact diagonalization solutions, both at and away from half filling.\cite {12}. 
Here we show that it provides accurate results also for nanostructures. 

While there is no exact result for multi-orbital many-atom systems available, as a proof of robustness of the approach we present results for the calculated magnetic properties  
 of ${\rm Fe}$ nanoparticles containing 11 to 100 atoms. We demonstrate that the results are in excellent agreement with experimental data,\cite{Knickelbein,Billas} and in most cases better than that obtained by DFT+U.\cite{Datta} The stage is now set for further application of this nano-DMFT+DFT approach.

{\it Method.}--Usually, the effects of strong electron-electron correlations are studied by solving a model described by the Hubbard Hamiltonian:   
\begin{eqnarray}
H=-\sum_{i,j,\sigma ,l,m}t_{il;jm}c_{i\sigma l}^{\dagger}c_{j\sigma m}
+\sum_{i,j,\sigma ,\sigma ',m}U_{i,j,\sigma, \sigma '}^{lm}n_{i\sigma l}n_{j\sigma ' m}.
\label{1}
\end{eqnarray}
where $c_{i\sigma l}^{\dagger}$ and $c_{i\sigma l}$  are the creation and annihilation operators of electron at site ${i}$ with spin $\sigma$ and other quantum number 
 (orbital energy level, orbital momentum etc) $l$; 
$c_{i\sigma l}^{\dagger}c_{i\sigma l}$=$n_{i\sigma l}$ is the particle number operator for state $\sigma , l$ at site $i$.
The kinetic energy and the potential energy (short-range Coulomb repulsion) are defined
by the hopping term, $t_{il;jm}$, and the Coulomb repulsion matrices, $U_{i,j,\sigma \sigma '}^{lm}$, respectively. 
The diagonal elements
$t_{il;il}$ define the energy of orbital $l$ at site $i$.
We obtain the hopping parameters directly from the 
calculations of the energy minimized (relaxed) structures of the system at hand using DFT or DFT+U.
For the Coulomb repulsion energy we invoke  the local approximation, $U_{i,j,\sigma \sigma '}^{lm}=\delta_{ij}\delta_{\sigma \sigma '}\delta_{lm}U^{l}$, since these terms usually contribute the most to the Hamiltonian in Eq.(\ref{1}),
and choose the values for $U^{l}$ ($l$ in this context being the orbital quantum number: s, p, d ...) as fitting parameters.
Note that one can in principle obtain the values for $U^{l}$ from DFT calculations.\cite{13} 
Note also that unsupported small clusters may
experience mechanical rotation, which may lead to effects related to the spin-mechanical orbital interaction, including a possibility of entanglement of corresponding degrees of freedom. Since study of such effects is beyond the scope of the article, we do not include the corresponding term into the Hamiltonian.

Many physical properties of the system described by the Hamiltonian in Eq.~(\ref{1}) can be obtained from the time-ordered Green's function (GF)
$
G_{i\sigma l;j\sigma ' m}(t,t')=-i\langle Tc_{i\sigma l}(t)c_{j\sigma ' m}^{\dagger}(t')\rangle .
$
In particular,  the site- and the orbital-spin densities that define the magnetic properties of the system
can be found from the following expression:
%\begin{eqnarray}
$n_{i,\sigma ,l}=-\int\frac{d \omega}{2\pi }{\rm Im} G_{i,\sigma ,l;i,\sigma ,l}(\omega)$.
In order to find the single-electron Green's function, one needs to find the corresponding self-energy of the electron
$\Sigma_{i\sigma l;j\sigma ' m}(\omega)$, which describes all effects of the electron-electron interaction.
The (inverse) time-ordered GF is related to the single-particle self-energy $\Sigma_{i\sigma l;j\sigma ' m}(t,t')$
via the Dyson equation (in frequency representation): 
\begin{eqnarray}
G_{i\sigma l;j\sigma ' m}^{-1}(\omega)
=
\delta_{\sigma ;\sigma ' }\delta_{ij}\delta_{lm}\omega -\delta_{\sigma ;\sigma ' }t_{il;jm}
-\Sigma_{i\sigma l;j\sigma ' m}(\omega).
\label{Dysonnonhom}
\end{eqnarray}
Thus, one can find the single-electron Green's function by inverting Eq. (\ref{Dysonnonhom}) provided that the self-energy is known.
The problem thus reduces to finding the electron self-energy for which we employ DMFT.

In the DMFT approximation for finite sized systems (N sites, M orbitals)
we adopt the generalization proposed by Nolting and Potthoff for non-homogeneous extended systems \cite{Potthoff1,Potthoff2} which assumes that the electronic self energy
is local in site and quantum indices, but depends on their value:
$
\Sigma_{i\sigma l;j\sigma ' m}(\omega)=\delta_{\sigma ;\sigma ' }\delta_{ij}\delta_{lm}
\Sigma_{i\sigma l}(\omega).
$
We thus need to find the GF matrix of size $2\times N\times L$ (2 stands for the number of spin degrees). 
As in the case of extended systems, the problem can be mapped onto the single-impurity problem.
Namely, in the DMFT approximation the many-site problem is reduced to the one-site problem
($2\times N\times L$ independent problems in the nonhomogeneous case) 
described by one-site one-orbital Hubbard Hamiltonian for an electron
in presence of a dynamical (time-, or frequency-, dependent) mean-field ${\cal G}_{i\sigma,l;i\sigma,l}(\omega )$.
This effective (bath) field describes the effect of the rest of the electrons in the nanostructure on the electron at site $i$ in orbital $l$.
This single-site (impurity) mapping means that the 
local GF $G_{i\sigma l;i\sigma l}(\omega)$  and the local self-energy $\Sigma_{i\sigma,l} (\omega))$
of the nanostructure are equal to the corresponding GF and self-energy of the impurity.
Provided ${\cal G}_{i\sigma,l;i\sigma,l}(\omega )$ is known one can find the impurity GF from
\begin{eqnarray} 
G_{i\sigma,l;i\sigma,l}(\omega )&=&\int D[\psi ] D[\psi^{*}]\psi_{i\sigma l}\psi_{i \sigma l}^{*}
\exp\left(
-\int_{0}^{\beta} d\tau
\right.
\nonumber \\
&\times&\left. 
\int_{0}^{\beta} d\tau '
%\right.
%\nonumber\\
%&\times&\left.
\sum_{\sigma }\psi_{i\sigma l}^{*}(\tau )
{\cal G}_{i \sigma l;i \sigma ,l}^{-1} (\tau -\tau ')
\psi_{i \sigma ,l}(\tau ')
\right.
\nonumber \\
&+&\left.
U_{i,l}\int_{0}^{\beta} d\tau n_{i\uparrow l}(\tau )n_{i\downarrow l}(\tau )
\right)  
\label{impuritynonhom}
\end{eqnarray}
($\psi^{*}$ and $\psi$ are here the Grassman fields\cite{grassman}, $\tau$ is the imaginary time variable, $\beta =1/T$ is the inverse temperature).\cite{6} The last equation is so called impurity problem equation.
On the other hand, to close the system of equations for local functions $G$, ${\cal G}$ and $\Sigma$ one can write down 
the impurity Dyson equation
\begin{eqnarray}
G_{i\sigma,l;i\sigma,l}(\omega )=1/({\cal G}_{i\sigma,l;i\sigma,l}^{-1}(\omega )-\Sigma_{i\sigma,l} (\omega)),
\label{Dysonnonhom2}
\end{eqnarray} 
which connects the single-site electron GF and self-energy.

In the case of finite sized systems one thus needs to solve the system of $2N\times M$ DMFT equations (\ref{Dysonnonhom}),  (\ref{impuritynonhom}) and (\ref{Dysonnonhom2}). 
One can do so by an iterative procedure for every value of frequency, for each site and orbital (see, e.g., Ref.~\cite{vturkowski}).
Since the equations can be solved separately
for every frequency, parallelization of the code is simple. However, 
non-homogeneity makes the calculations far more demanding, as
one needs to work with the inversion of rather large matrices (\ref{Dysonnonhom}). 
The most demanding part of the solution is the impurity problem  (\ref{impuritynonhom}).
While in principle, as for extended systems, the exact numerical solutions of this problem can be obtained
(by using, for example, Quantum Monte Carlo (QMC),\cite{6} or continuous-time QMC\cite{Gull} solvers),
in the case of multiple impurity problems it is much less computationally-demanding
to solve the problem the generalized  IPT approximation,\cite{12} developed for extended systems:
\begin{eqnarray}
\Sigma_{i,\sigma l}(\omega) 
= Un_{i\sigma l}+\frac{A_{i\sigma l}\Sigma_{i,\sigma l}^{(2)}(\omega)}{1-B_{i\sigma l}\Sigma_{i,\sigma l}^{(2)}(\omega)},
\label{IPT}
\end{eqnarray}
where $n_{i\sigma l}$ is the mean-field on-site orbital occupation ( of the HOMO,  LUMO and other atomic orbital states),
and $\Sigma_{i,\sigma l}^{(2)}(\omega)$ is the second-order self energy:
$
\Sigma_{i,\sigma l}^{(2)}(\omega_{n})=U^{2}\int_{0}^{\beta}d\tau {\rm exp} (i\omega_{n}\tau){\cal G}_{i,\sigma l}^{3}(\tau )
$
in the Matsubara frequency representation. One may easily obtain the corresponding expression
in the real frequency domain by applying analytical continuation.
The coefficients A and B are chosen to be:
$
A_{i\sigma l}=n_{i\sigma l}(1-n_{i\sigma l})/[n_{i\sigma l0}(1-n_{i\sigma l 0})], \ \
B_{i\sigma l}=[(1-2n_{i\sigma l})U_{i\sigma l}-\mu +{\tilde \mu}]/[n_{0i\sigma l}(1-n_{0i\sigma l})U_{i\sigma l}^{2}],
$
where $n_{i\sigma l0}=-(1/\pi )\int d\omega {\rm Im}{\cal G}_{i\sigma l;i\sigma l}(\omega )$.
The choice of A leads to the correct high-frequency behavior of the self-energy, while that of  B leads to the correct atomic limit
of this quantity. Above, we have introduced also parameters $\mu$ and ${\tilde \mu}$, chemical
potentials of the Green's functions $G$ and ${\cal G}$, similar to that for the extended case. 
There are different ways to 
choose the parameter ${\tilde \mu}$. One of the most reasonable is to use $n_{i\sigma l0}=n_{i\sigma l}$,
which leads to better agreement with the exact diagonalization result than $\mu={\tilde \mu}$.\cite{Helmes} Since for iron clusters we are far from the orbital  half-filling and the values of U are rather large,
to speed up the calculations we neglect the correction to the B parameter from the $\mu-{\tilde \mu}$ term and assume $\mu={\tilde \mu}$. In general, however, this term may lead to important effects.
It is important to note that the first (static) term on the right-hand side of Eq. (\ref{IPT}) corersponds to the
DMF+U (static mean-field) approximation.
The advantage of the IPT approximation comparing to the exact solvers
is its computational speed, which enables us to extend the calculation to system sizes that may otherwise not be possible.  Generally speaking, in the extended case IPT is regarded as a valid approximation small and large Us (including correct atomic limit). For intermediate values of the Coulomb repulsion IPT may be also regarded as a valid approximation as it reproduces important features, such as the central quasiparticle peak and the Hubbard bands in the DOS of extended systems,\cite{6} in agreement with more accurate solutions. 

{\it Application to Fe Nanoparticles.}-- The initial relaxed geometry for the $Fe_{19}$-$Fe_{100}$ clusters was chosen from the Cambridge Cluster Database \cite{Wales}, where the global minimum for iron clusters was obtained by using Finnis-Sinclair pair-potential\cite{FinnisSinclair} with initial bulk inter-atomic distances. To obtain smaller clusters we proceeded as follows. The initial structure for $Fe_{17}$ cluster was obtained by removing two atoms from all possible surface sites of the relaxed (icosahedral) $Fe_{19}$ cluster. Similarly, the structures $Fe_{15}$, $Fe_{13}$, and $Fe_{11}$ were obtained from the relaxed $Fe_{17}$, $Fe_{15}$, $Fe_{13}$ structures, correspondingly. 
For ionic relaxation we have used spin-polarized density functional theory (DFT) method as implemented in the Quantum Espresso code.\cite{Giannozzi} For the exchange and correlation potential the generalized gradient approximation (GGA) as parameterized by Perdew-Burke-Ernzerhof(PBE),\cite{Perdew} was used; ultra-soft pseudo-potentials were used for all atoms under consideration, the cut-off energy for the plane-wave expansion was 30Ry. No symmetry constraints were imposed on the structural geometry of clusters during structural relaxation. The structures were relaxed until the atomic forces converged to less than $0.01 eV/\AA$. We used a cubic simulation box for the relaxation of isolated freestanding clusters, with at least $12\AA$ distance between the cluster and its periodic images in all three directions. 
The resulting relaxed geometries of the clusters are presented in Fig. 1.
\begin{figure}[h]
\includegraphics[width=8.0cm]{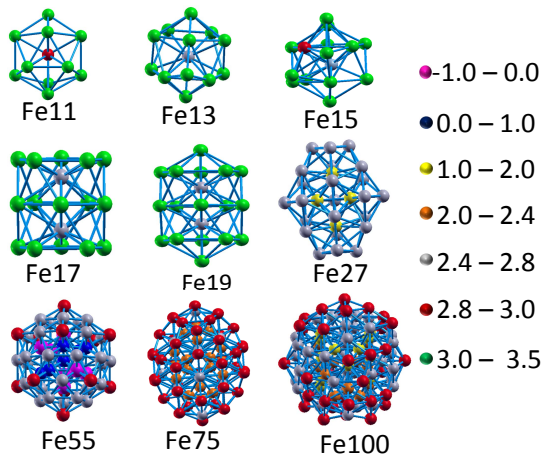}
\caption{\label{fig2} 
Relaxed structures of Fe clusters obtained with GGA+U. Different color indicate different values of magnetization(in unit of $\mu_{B}$) as indicated on the right.
}
\end{figure}

Clearly, the atoms on the surface have much higher magnetization than the atoms inside the cluster, and magnetization of the surface atoms increasing with decreasing cluster size. In one case ($Fe_{55}$), we find that some core atoms are anti-ferromagnetically coupled with the neighboring atoms(shown in pink). The fact that the atomic magnetization inside the cluster is lower compared to the surface atoms, can be explained by the higher coordination number of the bulk atoms, which increase the hybridization and hence the (itinerant type) ferromagnetism. The bondlength distribution of the relaxed clusters are presented in Fig.1 of the Supplementary Material.

The hopping parameters ${t}$ for the s- and d-valence electrons of the geometries presented in Fig.1 were obtained using the Slater-Koster matrix approximation\cite{23} (for details see Refs.~\cite{Turkowski,vturkowski}). With $t$ in hand (along with our already-chosen range of values for U), we now turn to DMFT, using a nano-DMFT code written by us (the particulars of which are described above).

Contrary to the standard DMFT codes for extended systems, which cannot be readily applied to systems with many atoms in the unit cell, our code can work with several hundreds atoms.
\begin{figure}[h]
\includegraphics[width=8.0cm]{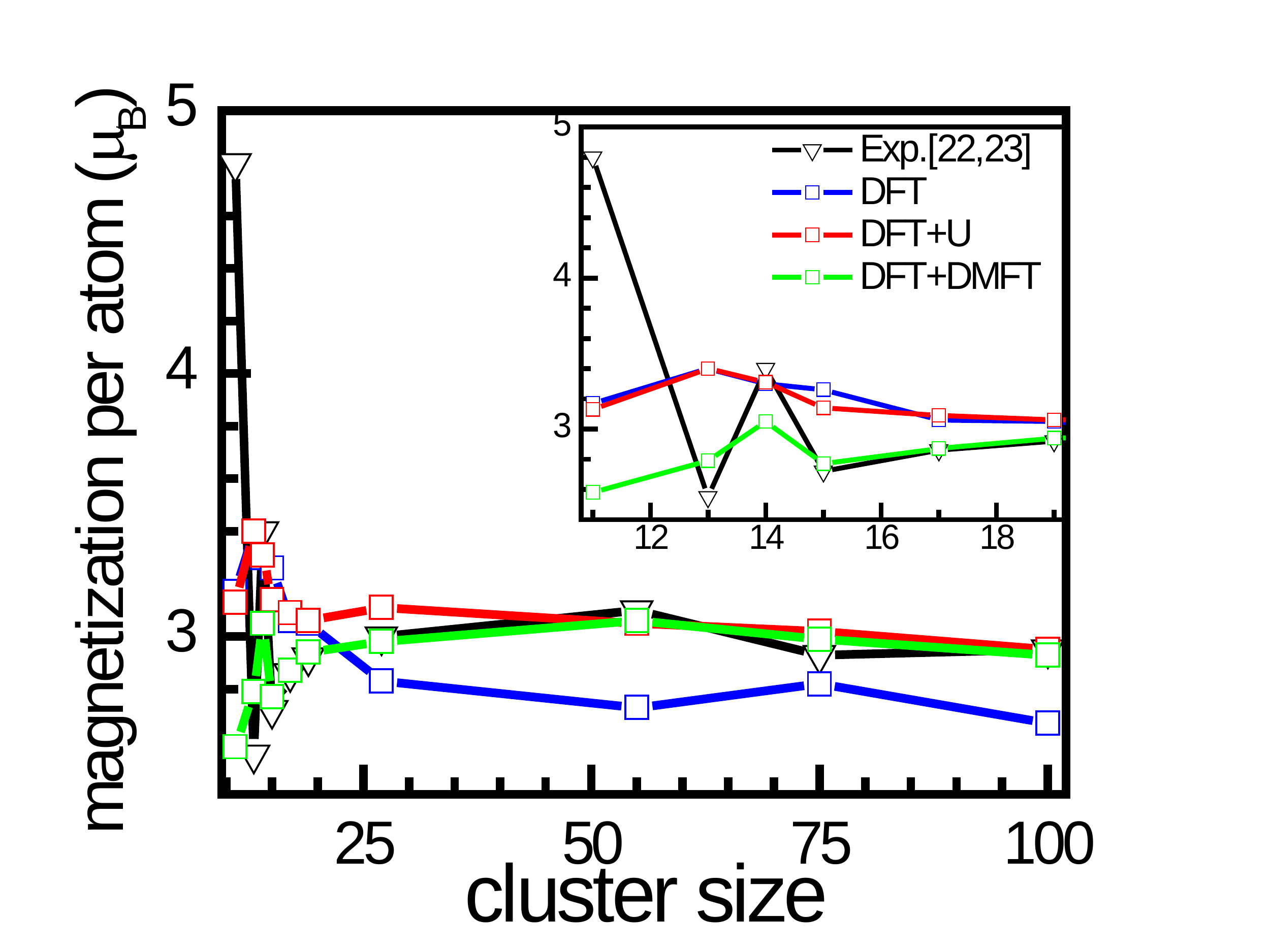}
\caption{\label{fig4} 
Magnetization per atom versus the cluster size in the case of  DFT, DFT+U, DMFT calculations and experimental data. On the inset figure detailed data for small clusters (up to 19 atoms) are shown. 
In the calculations, we have used U=2.3eV, in agreement with the estimations for the bulk case.
}
\end{figure}
As seen from Fig. 2, the DFT+U calculations tend to over-estimate magnetic moments, especially in small clusters, while inclusion of dynamical effects in general lead to a reduction of the magnetization. DMFT gives perfect agreement with experiment for all cases considered except that of  $Fe_{11}$. In order to allow the reader to make a more detailed comparison, we supplement Fig.2 with Table I, in which the values of the magnetization for different cases are presented.
\begin{table}
\caption{The magnetization per atom as function of the number of atoms in the Fe clusters at U=2.3eV. Along with the DFT+U and DFT+DMFT results we present a set of experimental data \cite{Knickelbein,Billas}.} \label{Table1}
\begin{ruledtabular}
\begin{tabular}{ccccc}
N     &DFT   & DFT+U                &  DFT+DMFT & exp.   \\
 11  &    3.17     &  3.13            &   2.58          &  4.8     \\
 13  &    3.40     &  3.40            &   2.79          &   2.55   \\
 14  &    3.30     &  3.31            &   3.05          &   3.40   \\
 15  &    3.26     &  3.14            &   2.77          &   2.72   \\
17   &    3.06     &  3.09            &   2.87         &    2.86   \\
19   &    3.05     &  3.06            &   2.94         &    2.92  \\
27   &   2.83     &   3.11            &   2.98         &    3.00   \\
55   &   2.73     &  3.05             &   3.06         &    3.10  \\
75   &   2.82     &  3.02             &   2.99         &    2.93  \\
100 &  2.67      &  2.95             &   2.93         &    2.95

\end{tabular}
\end{ruledtabular}
\end{table}

 In general such a  significant reduction of cluster magnetization may indicate that the orbital position (through static shift of the energy levels
resulting from self energy correction) and/or their occupancy (due to the frequency-dependence of the self-energy) may change dramatically when  dynamical effects are taken into account. A striking result in Figure 2 is the reproduction with DMFT of the local minimum for magnetization  of the 13-atom cluster.
Both DFT and DFT+U calculations fail in this regard. This minimum may  be related to the transition from an effective
"small cluster"- to bulk-like behavior, from low to large coordination number regime. 
As we know, when the number of atoms in a cluster is small, the number of under-coordinated atoms is large (surface atoms) and the system is more correlated (since the number of sites for electron to hop is small). Correlated systems tend to be antiferromagnetic, if the number of electrons is equal to that of the sites. With increasing number of atoms this strong correlations from a local exchange becomes less
pronounced.
On the other hand, with increasing number of atoms the system becomes more bulk-like and band (Stoner) ferromagnetism  becomes more dominant. Thus, with increasing nanoparticles size one initially has a decrease of magnetization due to destruction of the local exchange, followed by an increase from the build up of Stoner-type magnetism. The transition point N=13 (Fig.2) corresponds to the case in which the average coordination number becomes close to the bulk system.

We have also performed DFT+U and DFT+DMFT calculations for ${\rm  Fe_{11}-Fe_{19}}$ structures obtained from several different initial configurations.
 The trend for the magnetization is the same: DFT+DMFT significantly improves the results as compared to DFT+U  in all cases, except for ${\rm Fe_{11}}$ clusters for which none reproduces the high value extracted from experimental data. For details see Fig.2 in the Supplementary Material.

{\it Conclusions.}--We have proposed a DFT+nonhomogenous DMFT approach using the generalized IPT impurity solver to study the physical  properties of finite sized (non-extended) systems with strong electron-electron correlations.
Application of the approach to examine the magnetic moment of 11-100 atom Fe clusters shows that in most cases the inclusion of dynamical correlations results in better agreement with experiment \cite{Knickelbein,Billas} as it leads to a decrease in the calculated magnetization of the clusters.  This is to our knowledge the first demonstration that DMFT applied to experimentally realizable nanoparticles
can produce reliable results.
The methodology should have multiple applications as it can be readily applied to systems containing up to several hundred atoms and to other types of materials. The computational speed of the developed code is fast enough such that
the total computational time of the DFT+DMFT calculations is of the same order of magnitude as the corresponding time for the DFT calculations.
%There are several directions for future work. Most important of them 
%is the development of a reliable and more accurate impurity solver which will not significantly affect %the computational speed.
%For this purpose it would be helpful to first reproduce some rigorous results for strongly correlated %systems. For example, one can use the exact expressions for the electron GF and  self-energy %spectral moments.\cite{moments1,moments2,moments3} 

%Another direction is extension of the formalism to systems containing thousands
%of atoms. 
%In this case, one needs to look for a compromise between the system size and accuracy, due to %limitations on the real space Green's function matrix size. Some (possibly cluster or scaling) %approximations have to be developed for this case.
%Solving these problems will provide us with reliable and accurate tools to study electron correlation %effects in molecules and nanostructures of arbitrary size and composition.
 
{\it Acknowledgments.}--
We acknowledge DOE for financial support under Grant No, DOE Grant DE-FG02-07ER46354. We thank Lyman Baker for critical reading of the manuscript.

\end{document}